\documentclass[doublecol,figures]{epl2} 

\usepackage{amsmath,amssymb,mathtools}

\title{Cluster approximations for the TASEP: stationary state and dynamical transition}
\shorttitle{Cluster approximations for the TASEP} 

\author{A. Pelizzola\inst{1,2} and M. Pretti\inst{1,3}}
\shortauthor{A. Pelizzola \etal}

\institute{                    
  \inst{1} Dipartimento di Scienza Applicata e Tecnologia and CNISM,
  Politecnico di Torino, 
Corso Duca degli Abruzzi 24, I--10129 Torino, Italy\\
  \inst{2} INFN, Sezione di Torino, via Pietro Giuria 1, I-10125 Torino,
Italy\\
\inst{3} Consiglio Nazionale delle Ricerche-Istituto dei Sistemi Complessi (CNR--ISC)
}

\abstract{ We develop and test cluster approximations, which
  generalize simple mean--field by taking into account more and more
  local correlations, for the Totally Asymmetric Simple Exclusion
  Process with open boundaries. We consider in detail the pair and
  triplet approximations, discussing the improvements with respect to
  mean field in various steady state properties. Moreover, we
  analyze the recently discovered dynamical transition, describing how
  the spectrum of the relaxation matrix changes at the transition. }

\begin{document}

\maketitle

\section{Introduction}

In equilibrium statistical physics a huge research effort has been
devoted to develop approximations for models with a large number of
interacting variables, whose exact solution is typically out of reach
for both analytical and computational approaches. A large fraction of
this effort has been focused on mean field techniques and their
generalizations, yielding simple and efficient approaches, which
can turn out to be exact in certain limiting cases. 

In the last decades such an effort has been extended to nonequilibrium
statistical physics, in particular to the well defined and
interdisciplinarily relevant problem of models defined on graphs, whose
dynamics can be described in terms of a Markov process \cite{vanKampen},
paradigmatic examples being the kinetic Ising model and epidemic
processes.

A lot of techniques exist, which go beyond the simple mean--field
approach \cite{Kawasaki}. Here we briefly mention those which are most
relevant for the present work, for a more detailed review see
\cite{TransportBook}.
In many such approaches one assumes that the probability distribution
of a model with many variables factors into a suitable product of
local marginals, including of course mean--field as its simplest
version. This idea has led to the so--called cluster approximations
\cite{nm,Ito1995}, local structure theories
\cite{BoccaraBook,Knight1987}, $n$--step Markovian approximations
\cite{VulpianiBook}, and local equilibrium approximations
\cite{Kawasaki,AdvPhys,DeLos1,DeLos2,SchweitzerBehera2015}.
The Path Probability
Method (PPM) \cite{PPM0,PPM01,PPM1,PPM2,PPM3} is the dynamical version
of the well--known Cluster Variational Method \cite{CVM,An,MyRev} and,
like the latter, is based on the minimization of a suitable
free--energy--like functional obtained by a cluster expansion. It has
been shown to be related, in certain instances, to local equilibrium
approximations \cite{ZM-JSTAT}. The well--known cavity method
\cite{Mezard} has been recently extended to kinetic problems
\cite{DynCavity00,DynCavity0,DynCavity1,DynCavity2,DynCavity3,DMP1,DMP2}
yielding approaches known as dynamic cavity and dynamic message
passing \cite{Moore2015}, which have been shown to be efficient and
accurate on large systems. In the last few years, suitable
generalizations of the PPM idea have been shown
\cite{Pelizzola,RicciTersenghi,Pretti-SIS} to be often more efficient
and accurate than dynamic cavity and dynamic message passing. Although
based on a cluster expansion of a free--energy--like functional, these
approaches can in some instances be interpreted in terms of an
assumption of a suitable factorization of the model probability
distribution, as in local equilibrium approximations.

Some of the aforementioned approaches are not directly applicable to
another class of paradigmatic models in nonequilibrium statistical
physics, namely the Totally Asymmetric Simple Exclusion Process
(TASEP) and its extensions (see \cite{ChouMallickZia11} for a recent
review of these models and their applications). This can be understood
by considering the formulation of the dynamics. In both kinetic
Ising--like models and epidemic processes, a transition at a given
node (e.g.\ from susceptible to infected) can occur without implying a
transition on an adjacent node. In the TASEP we have particles moving on a
one--dimensional lattice, and a particle can hop to the adjacent node
provided it is empty, so the transition from occupied to empty at node
$i$ implies the transition from empty to occupied at node $i+1$.

In the present work we shall apply cluster approximations of the local
equilibrium type to the TASEP model. We shall explicitly consider the
pair and triplet approximations, corresponding to the $(2,1)$ and
$(3,2)$ approximations in \cite{nm}, or the ``triplet'' and
``quintuplet'' approximations in \cite{SchweitzerBehera2015}, which
take into account correlations between 2 and 3 adjacent nodes
respectively. In addition to discussing how these approximations can
improve the description of steady--state properties, we shall focus on
a recently discovered dynamical transition. In this case, before
applying the pair and triplet approximations, we shall investigate in
detail the mean field approximation, discussing in particular the
qualitative changes occurring at the transition in the spectrum of the
relaxation matrix.

\section{Model}
\label{sec:Model}

The TASEP is a paradigmatic model in nonequilibrium statistical
physics. It was originally introduced, in a more general form, in
\cite{MacDonaldGibbsPipkin68} as a model of mRNA translation. The
steady state of the TASEP with open boundaries \cite{Krug91} has been
exactly worked out in the 1990s
\cite{DerridaDomanyMukamel92,SchutzDomany93,Derrida-etal93,Derrida98},
showing a rich phase diagram, features which make it a suitable
candidate for testing approximate methods. The mean--field
approximation for the steady state was discussed in
\cite{DerridaDomanyMukamel92}. As previously mentioned, the model is
defined on a one--dimensional lattice,
where particles can hop only rightward, from each node to the adjacent
one, provided the latter is empty, at unit rate. 
Considering a lattice of length $N$, we label
lattice nodes from left to right by $i = 1, \ldots N$ and introduce
random variables of the occupation number type $n_i^t = 0, 1$, where
as customary $n_i^t = 1$ if node $i$ is occupied by a particle at time
$t$ and $n_i^t = 0$ otherwise. 
Expectation values will be denoted by $\langle \rangle$,
e.g.\ we will use the local densities $\rho_i^t = \langle n_i^t
\rangle$. Local densities in the stationary state do not depend on
time and will be denoted by $\rho_i$. Whenever the stationary density
profile exhibits a bulk region, we will denote the corresponding
density by $\rho$. 


Various types of boundary conditions have been considered in the
literature. Here we will focus on the particularly interesting case of
open boundary conditions, with particles injected at the leftmost node
(provided it is empty) with rate $\alpha$, and extracted from the
rightmost node (provided it is occupied) with rate $\beta$. Notice
that a fully equivalent description may be obtained assuming that the
system is in contact with two reservoirs of fixed densities $\alpha$
and $1 - \beta$, which respectively inject and extract particles at
unit rate. 

\begin{figure}
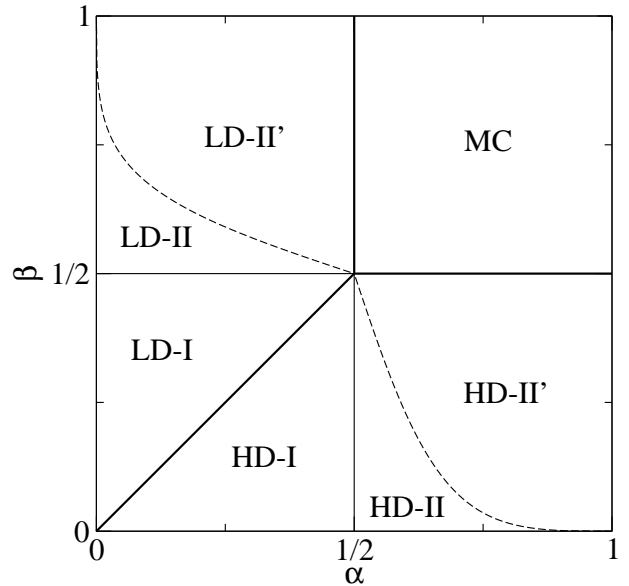

\onefigure[width=0.45\textwidth]{PhaseDiagramSquare.eps}
\caption{Phase diagram of the TASEP model with open boundary
  conditions. See main text for a description of the various phases
  and transitions between them.}
\label{fig:PhaseDiagram}
\end{figure}

Upon varying $\alpha$ and $\beta$ a rich, exactly known
\cite{DerridaDomanyMukamel92,SchutzDomany93,Derrida-etal93,Derrida98,deGierEssler05,deGierEssler08}
phase diagram, illustrated in Fig.\ \ref{fig:PhaseDiagram}, is
obtained.

If $\beta < 1/2$, $\alpha > \beta$, the steady state is in a {\em high
density phase}, with a current $J = \beta (1 - \beta)$ and a bulk
density $\rho = 1 - \beta > 1/2$, which extends up to the right
boundary. At the left boundary the density takes the value $1 -
J/\alpha$ and approaches the bulk with an exponential decay (in the
region denoted by HD-I in Fig.\ \ref{fig:PhaseDiagram}), with
power--law corrections for $\alpha \ge 1/2$ (regions denoted by HD-II
and HD-II' in Fig.\ \ref{fig:PhaseDiagram}).

Exploiting a particle--hole symmetry of the model, it can be
understood that for $\alpha < 1/2$, $\beta > \alpha$, the situation is
reversed: the steady state is in a {\em low density phase}, with $J =
\alpha (1 - \alpha)$ and $\rho = \alpha < 1/2$, which extends
up to the left boundary, while at the right boundary the density takes
the value $J/\beta$ and approaches the bulk with an exponential decay
(in the region denoted by LD-I in Fig.\ \ref{fig:PhaseDiagram}), with
power--law corrections for $\beta \ge 1/2$ (regions denoted by LD-II
and LD-II' in Fig.\ \ref{fig:PhaseDiagram}).

The above 2 phases are separated by the coexistence line $0 < \alpha =
\beta < 1/2$ (the thick line separating LD-I and HD-I regions in
Fig.\ \ref{fig:PhaseDiagram}), where the current is $J = \alpha (1 -
\alpha)$ and the densities at the left and right boundaries are
$\alpha$ and $1 - \alpha$ respectively, and are connected by a linear
density profile.

Finally, for $\alpha > 1/2$ and $\beta > 1/2$ the steady state is in a
{\em maximal current phase} (denoted by MC in
Fig.\ \ref{fig:PhaseDiagram}), with $J = 1/4$ and $\rho = 1/2$. The
densities at the left and right boundaries are $1 - J/\alpha > 1/2$
and $J/\beta < 1/2$ respectively, and the approach to the (central)
bulk value is power--law on both sides.

In addition to the above phases and transitions, a {\em dynamical phase
transition} (denoted by dashed lines in Fig.\ \ref{fig:PhaseDiagram}),
not corresponding to any steady--state transition, has recently
been discovered theoretically \cite{deGierEssler05,deGierEssler08}
(and numerically confirmed \cite{ProemeBlytheEvans11}) in the low and
high density phases. Considering the high density phase to fix ideas,
the transition can be characterized as follows: for any $\beta < 1/2$,
the (longest) relaxation time of the system is independent of $\alpha$
for $\alpha > \alpha_c(\beta)$ (the region denoted by HD-II' in
Fig.\ \ref{fig:PhaseDiagram}), where
\begin{equation}
\label{eq:dGEline}
\alpha_c(\beta) = \left[ 1 + \left( \frac{\beta}{1-\beta}
  \right)^{1/3} \right]^{-1},
\end{equation}
while for $\alpha < \alpha_c(\beta)$ the relaxation time depends on
both $\alpha$ and $\beta$. 


\section{Cluster approximations}

Simple mean--field--like, cluster--based approximations can be obtained
by assuming that, at a given time $t$, the model probability distribution
factors as a suitable product of local marginals. 

To be more quantitative, it is useful to define a
class of marginals, involving a string of $k + 1$ (with $k \ge 0$)
adjacent nodes: $P_i^t[n_i n_{i+1} \cdots n_{i+k}]$ will denote the
probability that, at time $t$, the occupation numbers of nodes from
$i$ to $i+k$ take values $n_i, n_{i+1}, \cdots n_{i+k}$
respectively. In order to model boundary conditions we will introduce
two auxiliary nodes $i = 0$ and $i = N+1$ of fixed densities $\rho_0 =
\alpha$ and $\rho_{N+1} = 1 - \beta$ respectively, and we will assume
\begin{equation}
  {P}_{0}^{t}[1 n_1 \dots n_i]
  \equiv
  \alpha {P}_{1}^{t}[n_1 \dots n_i], \qquad i = 1, \ldots, N
\end{equation}
and
\begin{equation}
  {P}_{i}^{t}[n_i \dots n_N 0]
  \equiv
  \beta {P}_{i}^{t}[n_i \dots n_N], \qquad i = 1, \ldots, N. 
\end{equation}

It is immediate to write exact dynamical evolution equations for a few
of the above marginals. The time evolution of the local densities
$\rho_i^t = P_i^t[1]$ is given by
\begin{equation}
  \dot{P}_{i}^{t}[1]
  = {P}_{i-1}^{t}[10]
  - {P}_{i}^{t}[10], \qquad i = 1, \ldots, N,
\label{eq:LocalDensityTimeEvo}
\end{equation}
where the local currents $J_i^t = P_i^t[10]$ appear. The time
evolution of the local currents is in turn given by
\begin{equation}
  \dot{P}_{i}^{t}[10]
  = {P}_{i-1}^{t}[100]
  - {P}_{i}^{t}[10]
  + {P}_{i}^{t}[110], \quad i = 1, \ldots, N-1,
\label{eq:LocalCurrentTimeEvo}
\end{equation}
where certain 3--node marginals appear. The time evolution of these
3--node marginals is then given by
\begin{alignat}{3}
  \dot{P}_{i}^{t}[100] &
  = {P}_{i-1}^{t}[1000]
  - {P}_{i}^{t}[100]
  + {P}_{i}^{t}[1010], \nonumber
  \\
  \dot{P}_{i}^{t}[110] &
  = {P}_{i-1}^{t}[1010]
  - {P}_{i}^{t}[110]
  + {P}_{i}^{t}[1110]
\label{eq:3NodeTimeEvo}
\end{alignat}
($i = 1, \ldots, N-2$) and involves 4--node marginals, and so
on. Clearly, in order to close our set of equations, we need to
introduce approximations. 

The simplest possible choice is the ordinary mean--field approximation
\cite{DerridaDomanyMukamel92}, where one assumes that
\begin{equation}
\label{eq:MeanField}
P_i^t[n_in_{i+1}] = P_i^t[n_i] P_{i+1}^t[n_{i+1}]. 
\end{equation}
Using this assumption in Eq.\ \ref{eq:LocalDensityTimeEvo} we obtain a
set of closed equations for the local densities:
\begin{equation}
\label{eq:LocalDensityTimeEvoMF}
\dot\rho_i^t = \begin{cases}
\rho_{i-1}^t (1 - \rho_i^t) -  \rho_i^t (1 - \rho_{i+1}^t), & i = 2
\ldots, N-1 \\
\alpha (1 - \rho_1^t) -  \rho_1^t (1 - \rho_2^t),  & i = 1 \\
\rho_{N-1}^t (1 - \rho_N^t) - \beta \rho_N^t, & i = N.
\end{cases}
\end{equation}

\subsection{Pair approximation}

A second, more accurate, level of approximation is obtained by
retaining correlations between adjacent nodes, that is by
assuming
\begin{equation}
\label{eq:Pair}
P_i^t[n_i n_{i+1} n_{i+2}] = \frac{P_i^t[n_i n_{i+1}]
  P_{i+1}^t[n_{i+1} n_{i+2}]}{P_{i+1}^t[n_{i+1}]}
\end{equation}
as in the $(2,1)$ approximation in \cite{nm} or in the ``triplet''
approximation in \cite{SchweitzerBehera2015}.
In order to obtain a set of dynamical evolution equations for the $2 N
- 1$ variables $\{\rho_i^t, J_i^t\}$, we first observe that
Eq.\ \ref{eq:LocalDensityTimeEvo} can
be rewritten, with no need of approximations, as
\begin{equation}
\label{eq:LocalDensityTimeEvoPair}
\dot\rho_i^t = \begin{cases}
J_{i-1}^t -  J_i^t, & i = 2, \ldots, N-1 \\
\alpha (1 - \rho_1^t) -  J_1^t, & i = 1 \\
J_{N-1}^t - \beta \rho_N^t, & i = N.
\end{cases}
\end{equation}
In addition, 
we close Eq.\ \ref{eq:LocalCurrentTimeEvo} using the
approximation Eq.\ \ref{eq:Pair}, obtaining
\begin{equation}
\label{eq:LocalCurrentTimeEvoPair}
\dot J_i^t = \frac{J_{i-1}^t(1 - \rho_{i+1}^t - J_i^t)}{1-\rho_i^t} -
J_i^t + \frac{(\rho_i^t - J_i^t) J_{i+1}^t}{\rho_{i+1}^t}
\end{equation}
for $i = 2, \ldots, N-1$, 
\begin{equation}
\label{eq:LocalCurrentTimeEvoPair1}
\dot J_1^t = \alpha (1 - \rho_{2}^t - J_1^t) - J_1^t + \frac{(\rho_1^t
  - J_1^t) J_{2}^t}{\rho_{2}^t},
\end{equation}
and 
\begin{equation}
\label{eq:LocalCurrentTimeEvoPairN}
\dot J_{N-1}^t = \frac{J_{N-2}^t(1 - \rho_{N}^t -
  J_{N-1}^t)}{1-\rho_{N-1}^t} - J_{N-1}^t + \beta (\rho_{N-1}^t -
J_{N-1}^t) .
\end{equation}

\subsection{Triplet approximation}

A third level of approximation is obtained by retaining correlations
between 3 adjacent nodes, that is by assuming
\begin{equation}
\label{eq:Triplet}
P_i^t[n_i n_{i+1} n_{i+2} n_{i+3}] = \frac{P_i^t[n_i n_{i+1} n_{i+2}]
  P_{i+1}^t[n_{i+1} n_{i+2} n_{i+3}]}{P_{i+1}^t[n_{i+1} n_{i+2}]}
\end{equation}
as in the $(3,2)$ approximation in \cite{nm} or in the ``quintuplet''
approximation in \cite{SchweitzerBehera2015}. Using the above
assumption in Eqs.\ \ref{eq:3NodeTimeEvo} we obtain equations which,
together with Eqs.\ \ref{eq:LocalDensityTimeEvo} and
\ref{eq:LocalCurrentTimeEvo},
form a closed set, involving $4N-5$ variables. Calculations at the
triplet level are straightforward, though cumbersome, extensions of
the pair approximation ones. In the following, in order to keep our
presentation contained, we will mainly focus on the pair
approximation, but we will comment on the main changes at the triplet
level and of course present the corresponding results. 

\subsection{Steady state in the pair and triplet approximations}

The equations derived above for the pair approximation, in particular
Eqs.\ \ref{eq:LocalDensityTimeEvoPair} and
\ref{eq:LocalCurrentTimeEvoPair}--\ref{eq:LocalCurrentTimeEvoPairN},
can be directly used to describe the 
steady state of the model. In the steady state, the variables
$\{\rho_i^t, J_i^t\}$ become time--independent. From the continuity
equation Eq.\ \ref{eq:LocalDensityTimeEvoPair} we obtain 
\begin{equation}
\label{eq:Continuity}
J_i = J_{i-1}, \qquad i = 2, \ldots, N-1,
\end{equation}
meaning that the stationary current is site--independent (so that we 
shall denote it by $J = J_i$, $i = 1, \ldots N-1$), 
together with the boundary conditions
\begin{equation}
\label{eq:Boundary}
\rho_1 = 1 - \frac{J}{\alpha}, \qquad \rho_N = \frac{J}{\beta}.
\end{equation}
In addition, from
Eqs.\ \ref{eq:LocalCurrentTimeEvoPair}--\ref{eq:LocalCurrentTimeEvoPairN}
and using the above boundary conditions (which happen to be exact), we
obtain the (approximate) recursive equation
\begin{equation} 
\label{eq:Recursion}
\rho_{i+1} (\rho_{i+1} - \rho_i + J) = (1 - \rho_i) (\rho_i - J),
\end{equation}
which holds for $i = 1, \ldots, N-1$ and can be solved
w.r.t.\ $\rho_{i+1}$, obtaining 
\begin{equation}
\label{eq:FwdRecursion}
\rho_{i+1} = \frac{\rho_i - J}{2} \left( 1 + \sqrt{1 +
    4 \frac{1 - \rho_i}{\rho_i - J}} \right).
\end{equation}
Of course, one could also solve w.r.t.\ $\rho_i$ and obtain the
inverse of the above recursion. These recursions have the same fixed
points 
\begin{equation}
\label{eq:FixedPoints}
\rho_\pm = \frac{1 \pm \sqrt{1 - 4 J}}{2}
\end{equation}
as in mean--field, and hence the bulk density $\rho$ and the currrent
$J$ satisfy the exact relation $J = \rho(1-\rho)$. 

A numerical solution of Eq.\ \ref{eq:FwdRecursion} with the boundary
conditions Eq.\ \ref{eq:Boundary} yields the density profiles in the
various phases of the model. The phase diagram turns out to coincide
with the mean--field one (which is known to be exact), but of course
one can obtain better approximations for the density profiles.  In
particular, in the low--density phase ($\alpha < 1/2$, $\beta >
\alpha$, bulk density $\rho = \alpha < 1/2$, boundary densities
$\rho_1 = \alpha$ and $\rho_N = \alpha (1 - \alpha)/\beta$, and
current $J = \alpha (1 - \alpha) < 1/4$), the approach to the bulk
density (from the right) is exponential, as in mean--field, but with a
better approximation for the characteristic length. Setting $x_i =
\rho_{i} - \rho$ we find $x_{i+1} = \gamma x_i$, where
\begin{equation}
\label{eq:Length}
\gamma \equiv \gamma(\rho) = \frac{1 - \rho^2}{\rho (2
  - \rho)},
\end{equation}
corresponding to a characteristic length $\xi$ given by
\begin{equation}
\label{eq:xiLD}
\xi^{-1} = \ln \gamma(\alpha).
\end{equation}
In the triplet approximation the approach to the bulk is described in
terms of $x_i$ and the 2 additional quantities
\begin{eqnarray}
y_i &=& P_i[100] - \rho (1-\rho)^2, \nonumber \\
z_i &=& P_i[110] - \rho^2 (1-\rho), \nonumber
\end{eqnarray}
where the superscript denoting time has been dropped in the stationary
state marginals. Linearizing close to the bulk we obtain
\begin{equation}
\left(
  \begin{array}{l}
    {x}_{i+1} \\
    {y}_{i} \\
    {z}_{i} \\
  \end{array}
\right)
=
\Gamma
\left(
  \begin{array}{l}
    {x}_{i} \\
    {y}_{i-1} \\
    {z}_{i-1} \\
  \end{array}
\right),
\end{equation}
where
\begin{equation}
\Gamma(\rho) = \textstyle \frac{1}{\rho(2-\rho)} \left(
\begin{array}{rrr}
  \scriptstyle 2\rho(1-\rho) & \scriptstyle -(2-\rho) & \scriptstyle -(1-\rho) \\
  \scriptstyle 2-\rho[5-3\rho(2-\rho)] & \scriptstyle (2-\rho)(3-\rho) & \scriptstyle 2(1-\rho)^2 \\
  \scriptstyle 0 & \scriptstyle -\rho(2-\rho) & \scriptstyle 0
\end{array}
\right),
\end{equation}
from which the inverse characteristic length $\xi^{-1}$ can be
obtained as the logarithm of the second smallest eigenvalue of
$\Gamma$. The smallest eigenvalue turns out to be less than 1, which
corresponds to an attractive direction of the low--density fixed
point, but one can verify that any non--zero contribution along this
direction is forbidden by the boundary conditions on the right side.
This implies that some care is needed when solving numerically the
equations for the density profile.

\begin{figure}
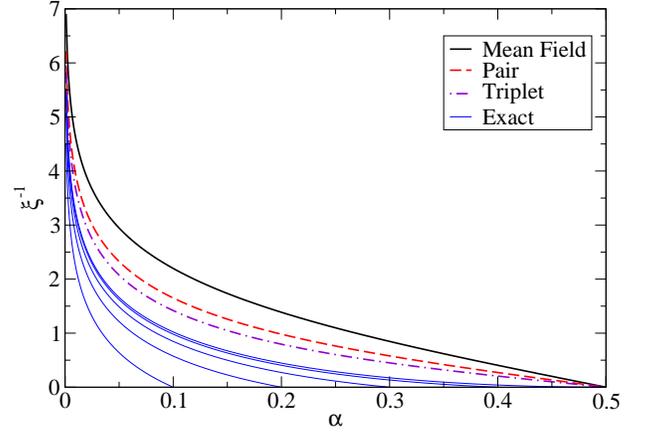

\onefigure[width=0.45\textwidth]{invxiLD.eps}
\caption{Inverse characteristic length vs.\ $\alpha$ in the
  low--density phase. From top to bottom: mean--field approximation
  (full line), pair approximation (dashed), triplet (dot--dashed),
  exact results (thin full lines) for $\beta > 1/2$, $\beta = 0.4,
  0.3, 0.2, 0.1$.}
\label{fig:xiLD}
\end{figure}

The (inverse) characteristic lengths estimated by the pair and triplet
approximations are plotted in Fig.\ \ref{fig:xiLD} and compared with
the mean--field result $\xi^{-1} = \ln \frac{1-\alpha}{\alpha}$ and
with the exact result $\xi^{-1} = \xi_\alpha^{-1} - \xi_\beta^{-1}$
(for $\beta < 1/2$) or $\xi^{-1} = \xi_\alpha^{-1}$ (for $\beta >
1/2$), where $\xi_\sigma^{-1} = - \ln \left[ 4 \sigma (1 - \sigma)
  \right]$ for $\sigma = \alpha, \beta$. Notice that the pair and triplet
approximations, as well as mean field, are not able to reproduce the
$\beta$ dependence of the characteristic length for $\beta < 1/2$.

\begin{figure}
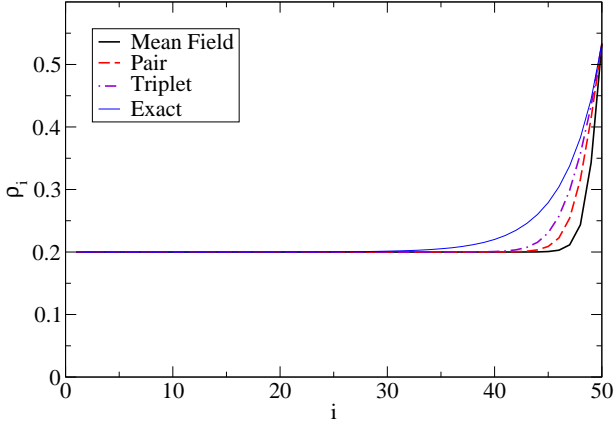

\onefigure[width=0.45\textwidth]{LDphase-a02b03-N50.eps}
\caption{Density profiles for $N = 50$, $\alpha = 0.2$, $\beta = 0.3$,
  (low--density phase). From bottom to top: mean
  field (full line), pair approximation (dashed), triplet
  (dot--dashed), exact result (thin full line).}
\label{fig:LDprofiles}
\end{figure}

In Fig.\ \ref{fig:LDprofiles} we report the density profiles, and
compare them to the mean field and exact ones, for $N = 50$, $\alpha =
0.2$ and $\beta = 0.3$.

We do not show the equivalent of
Figs.\ \ref{fig:xiLD}--\ref{fig:LDprofiles} in the high--density phase
since they would be simply qualitatively symmetrical, due to the
aforementioned particle--hole symmetry of the model, which is
preserved by the approximate theory. We just mention
that the characteristic length is given by
\begin{equation}
\label{eq:xiHD}
\xi^{-1} = -\ln \gamma(1 - \beta) = \ln \gamma(\beta),
\end{equation}
where the latter equality easily follows from Eq.\ \ref{eq:Length}.

The two low-- and high--density phases are separated by the
coexistence line $\alpha = \beta < 1/2$. The density profile is
symmetric and in the large $N$ limit it exhibits 2 macroscopic
(``bulk'') regions, separated by a domain wall: a low--density one,
with density $\rho_- = \alpha$, extending to the left boundary, and a
high--density one, with density $\rho_+ = 1 - \beta$, extending to the
right. The current is $J = \alpha (1 - \alpha) = \beta (1 - \beta)$
and the approach to both macroscopic regions is exponential, with a
unique characteristic length which can be obtained by
Eq.\ \ref{eq:xiLD} or, equivalently, Eq.\ \ref{eq:xiHD}. In the limit
$N \to \infty$ the domain wall can be anywhere and an infinite number
of solutions appear. This result is the same as in the mean field
approximation, while the exact solution averages over all the
positions of the domain wall, yielding a linear density profile.

\begin{figure}
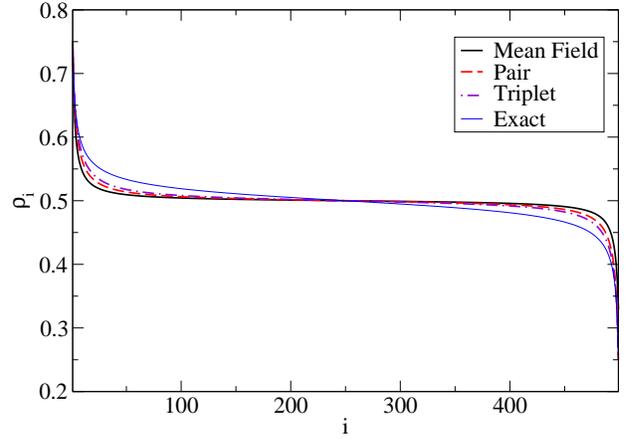

\onefigure[width=0.45\textwidth]{MCphase-a1b1-N499.eps}
\caption{Density profiles for $N = 499$, $\alpha = \beta = 1$,
  corresponding to the maximal current phase. Mean
  field (full line), pair approximation (dashed), triplet
  (dot--dashed), exact result (thin full line).}
\label{fig:MCprofiles}
\end{figure}

Finally, for $\alpha > 1/2$ and $\beta > 1/2$ we recover the
{\bf maximal current} phase, with bulk density $\rho = 1/2$, current
$J = 1/4$ and boundary densities $\rho_1 = 1 - 1/(4 \alpha) > 1/2$ and
$\rho_N = 1/(4\beta) < 1/2$. In order to understand how the bulk
density is approached from both sides, we need to expand
Eq.\ \ref{eq:Recursion} to 2nd order in $x_i, x_{i+1}$, obtaining
\begin{equation}
\label{eq:2ndOrderRecursion}
x_{i+1} - x_i \simeq - \frac{4}{3} x_i^2,
\end{equation}
which is solved by $x_i \sim i^{-1}$ (or, from the right, $x_i \sim
(N-i)^{-1}$). This $i^{-1}$ behaviour, also given by the triplet
approximation, is qualitatively the same as in the mean field
approximation, while the exact solution gives also a power law, but
with an exponent $1/2$ instead of 1.  In Fig.\ \ref{fig:MCprofiles} we
report the density profiles in the pair and triplet approximations,
and compare them to the mean field and exact ones, for $N = 499$
(using odd $N$ is technically convenient because it allows to start
the recursions from $\rho_{(N+1)/2} = 1/2$) and $\alpha = \beta =
1$. Due to the finite size, the current is slightly larger than $1/4$,
in particular $J \simeq 0.2500218$ in the pair approximation and $J
\simeq 0.2500380$ in the triplet approximation, to be compared with
the exact result 0.2507508 and the mean field one 0.2500097.

\subsection{Dynamical transition in the mean field approximation}

The approach to the steady state can be quantitatively studied by
linearizing the mean field dynamical equations,
Eq.\ \ref{eq:LocalDensityTimeEvoMF}, close to their steady state fixed
point. Denoting by $\rho_i$ the steady state value of $\rho_i^t$ and
linearizing we have 
\begin{equation}
\label{eq:LinearizationMF}
\dot\rho_i^t = - \sum_{j = 1}^{N}M_{ij} (\rho_j^t - \rho_j),
\qquad i = 1, \ldots, N.
\end{equation}
The matrix $M$ characterizes relaxation. In particular, its smallest
eigenvalue is the slowest relaxation rate, that is the inverse of the
longest relaxation time. The numerical computation of its spectrum is
particularly easy, since $M$ is tridiagonal with elements
\begin{eqnarray}
\label{eq:M-elements}
M_{i,i} &=& \frac{J}{\rho_i(1 - \rho_i)}, \qquad i = 1, \ldots, N,
\nonumber \\
M_{i,i+1} &=& - \frac{J}{1 - \rho_{i+1}}, \qquad i = 1, \ldots, N-1,
\nonumber \\
M_{i,i-1} &=& - \frac{J}{\rho_{i-1}}, \qquad i = 2, \ldots, N.
\end{eqnarray}
Considering the high--density phase to fix ideas, we observe that in
the large $N$ limit, except for its small $i$ (i.e.\ top--left)
portion, $M$ takes the Toeplitz form
\begin{equation}
\label{eq:M-Toeplitz}
M = \begin{pmatrix*}[l]
\ddots & \ddots & \\
\ddots & \ddots & - (1 - \beta) \\
& -\beta & 1
\end{pmatrix*}.
\end{equation}
The eigenvalues and eigenvectors of a tridiagonal Toeplitz matrix are
well--known (see e.g.\ \cite{Smith}). With the elements in
Eq.\ \ref{eq:M-Toeplitz} the eigenvalues are given by
\begin{equation}
\label{eq:M-eigenvalues}
\lambda_k = 1 - 2 \sqrt{\beta(1 - \beta)} \cos\frac{k\pi}{N+1} = 
1 - 2 \sqrt{J} \cos\frac{k\pi}{N+1}, 
\end{equation}
for $k = 1, \ldots, N$, and the smallest one tends to
\begin{equation}
\label{eq:M-smallest}
\lambda_1 = 1 - 2 \sqrt{\beta(1 - \beta)} = 1 - 2 \sqrt{J}. 
\end{equation}
This spectrum is functionally similar (though with
different coefficients) to that given by the Domain Wall Theory (DWT)
\cite{DWT1,DWT2}. 

\begin{figure}
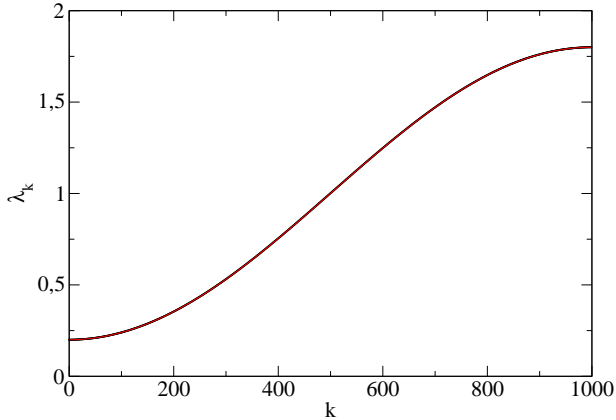

\onefigure[width=0.45\textwidth]{MFvsToeplitzSpectrum-a1b02-N1000.eps}
\caption{Eigenvalues of the mean--field relaxation matrix (circles),
  compared with the eigenvalues of the corresponding Toeplitz matrix
  (line), for $\alpha = 1$, $\beta = 0.2$ and $N = 1000$.}
\label{fig:MFvsToeplitzSpectrum}
\end{figure}

\begin{figure}
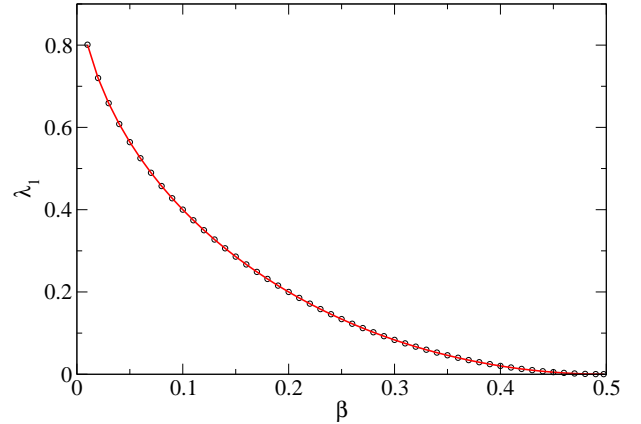

\onefigure[width=0.45\textwidth]{MFvsToeplitzRate-a1-N200.eps}
\caption{Smallest eigenvalue of the mean--field relaxation matrix
  (black circles), compared with that of the corresponding Toeplitz
  matrix (red line), Eq.\ \protect\ref{eq:M-smallest} for $\alpha = 1$
  and $N = 200$.}
\label{fig:MFvsToeplitzRate}
\end{figure}

One could wonder whether, at least in some cases, the spectrum of
$M$ is dominated by its Toeplitz--like macroscopic portion. Indeed,
this is what we find for $\alpha$ larger than a certain threshold
$\alpha_c(\beta)$ (to be determined later), which defines
a region in the mean--field phase diagram corresponding to HD-II' in
Fig.\ \ref{fig:PhaseDiagram}. In Fig.\ \ref{fig:MFvsToeplitzSpectrum}
we report the (numerically evaluated) eigenvalues of the mean field
relaxation matrix Eq.\ \ref{eq:M-elements} (black circles), together
with the eigenvalues (Eq.\ \ref{eq:M-eigenvalues}) of the
corresponding Toeplitz matrix (red line), for $\alpha = 1$, $\beta =
0.2$ and $N = 1000$. No differences are seen on the scale of the
figure, the largest (absolute) difference is slightly less than $4
\cdot 10^{-4}$. Moreover, in Fig.\ \ref{fig:MFvsToeplitzRate} we
report the (numerically evaluated) smallest eigenvalue $\lambda_1$ of
the mean field relaxation matrix (black circles), together with the
analytical estimate Eq.\ \ref{eq:M-smallest} (red line) for $\alpha =
1$ and $N = 200$, as a function of $\beta$. Again, we see perfect
agreement.  Notice that the analytical estimate
Eq.\ \ref{eq:M-smallest} tends to the result given by the Burgers
equation \cite{ProemeBlytheEvans11} approximation as $\beta$
approaches 1/2, but in general is different from this result.

The above situation changes qualitatively for $\alpha \in
(\beta,\alpha_c(\beta))$, that is in a region of the mean--field phase
diagram corresponding to HD-I and HD-II in
Fig.\ \ref{fig:PhaseDiagram}. In addition to a ``band'' of $N-1$
Toeplitz--like eigenvalues $\lambda_k \in (1 - 2\sqrt{J}, 1 + 2
\sqrt{J}), k = 2, \ldots N$, distributed with the same density as for
$\alpha > \alpha_c(\beta)$, we have that the smallest eigenvalue
$\lambda_1 < 1 - 2\sqrt{J}$, which gives the slowest relaxation rate,
detaches from the others. See Fig.\ \ref{fig:Rate-beta02} for the
behaviour of the smallest eigenvalue in the case $\beta = 0.2$,
compared to the pair and triplet approximations (see below) and the
exact and DWT results.

In Fig.\ \ref{fig:Eigenvectors} we plot the components $v_i$ of the
eigenvector corresponding to the smallest eigenvalue for $\beta =
0.2$, various values of $\alpha$ in the two regions ($\alpha_c(\beta =
0.2) \simeq 0.55$ in mean field approximation), and $N = 1000$. It can
be observed that the slowest relaxation mode is confined to the left
boundary, the more so the closer $\alpha$ to the dynamical transition,
but no qualitative difference can be seen between region HD-II, where
the rate depends on $\alpha$, and HD-II', where the rate is
independent of $\alpha$.

\begin{figure}
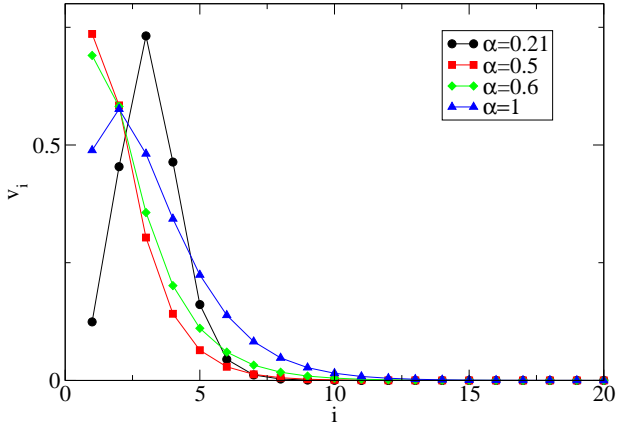

\onefigure[width=0.45\textwidth]{Eigenvectors.eps}
\caption{Components $v_i$ of the eigenvector corresponding to the
  smallest eigenvalue for $N = 1000$, $\beta = 0.2$ and $\alpha =
  0.21$  (circles), 0.5 (squares), 0.6 (diamonds), 1 (triangles).}
\label{fig:Eigenvectors}
\end{figure}

The line $\alpha_c(\beta)$ separating the two regions (given, in
the large $N$ limit, by Eq.\ \ref{eq:M-smallest}) is reported in
Fig.\ \ref{fig:DynTrans} for $N = 800$ and compared with the pair and
triplet approximations and exact and DWT results ($N = \infty$).

\subsection{Dynamical transition in the pair and triplet approximations}

The dynamical evolution equations characterizing the pair
approximation,
Eqs.\ \ref{eq:LocalDensityTimeEvoPair}--\ref{eq:LocalCurrentTimeEvoPairN},
can be conveniently rewritten in matrix form by introducing the
($2N-1$--component) vector $x^t = (x_1^t, \ldots, x_{2N-1}^t) =
(\rho_1^t, J_1^t, \cdots, \rho_{N-1}^t, J_{N-1}^t, \rho_N^t)$. Denoting
by $x_i$ the steady state value of $x_i^t$, and linearizing around
$x_i$, we have
\begin{equation}
\label{eq:Linearization}
\dot x_i^t = - \sum_{j = 1}^{2N-1}Q_{ij} (x_j^t - x_j),
\qquad i = 1, \ldots, 2N - 1,
\end{equation}
where $Q$ is a pentadiagonal 2--Toeplitz \cite{GoverBarnett} matrix
whose eigenvalues have a positive real part. At odds with mean field,
in this case we find that there are several pairs of complex conjugate
eigenvalues. The smallest eigenvalue is real and is again the
(slowest) relaxation rate $\lambda$ or, equivalently, the inverse of
the (longest) relaxation time of the dynamics. Unfortunately, we are
not aware of any exact result for the spectrum of such a
matrix. However we observe, on a purely numerical basis, that for $\alpha >
\alpha_c(\beta)$ the ratio between the mean field rate and the pair
one is nearly constant and very close to $\sqrt{2}$. In the case of
the triplet approximation the matrix to consider is of course of order
$4N-5$ and the results are qualitatively similar. 

\begin{figure}
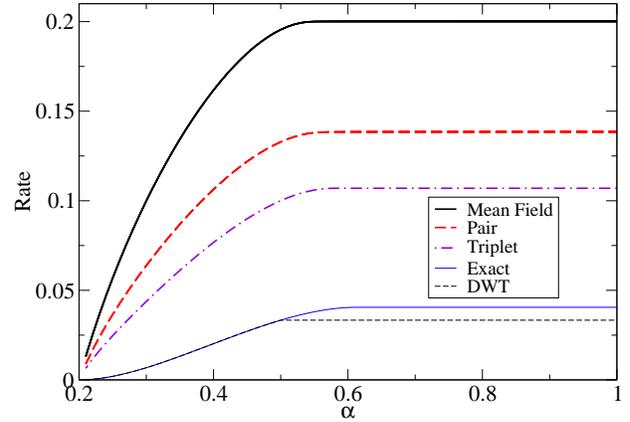

\onefigure[width=0.45\textwidth]{Rate-beta02-Coarse.eps}
\caption{Relaxation rate $\lambda$ for $\beta = 0.2$ as a function of
  the injection rate $\alpha$. From top to bottom: mean field (full
  line), pair approximation (dashed), triplet (dot--dashed), exact
  (thin full line) and DWT (thin dashed line) results. For the mean
  field, pair and triplet approximations, results for $N = 100, 200,
  400, 800$ are reported, but the differences are smaller than the
  line thickness.}
\label{fig:Rate-beta02}
\end{figure}

In Fig.\ \ref{fig:Rate-beta02} we report the relaxation rate $\lambda$
for $\beta = 0.2$ estimated by the pair and triplet approximation,
together with the results from the mean field approximation, the DWT
and the exact result. It can be observed that the pair
and triplet approximations improve over the mean field one,
considering both the rate value and the dynamical
transition point $\alpha_c(\beta)$. The properties of the spectrum are
similar to the mean field ones, that is for $\alpha > \alpha_c(\beta)$
the real parts of the eigenvalues form a ``band'' which tends to be
continuous in a finite interval as $N$ gets large, while for $\alpha
\in (\beta, \alpha_c(\beta))$, the smallest eigenvalue detaches from
the others and becomes $\alpha$--dependent. Comparing with DWT results
we see that mean field, pair and triplet overstimate the rate while
DWT underestimates it (for $\alpha > 1/2$), and that even the triplet
approximation does not reach the same accuracy as DWT. It is however
interesting to note that the pair and triplet approximations provide a
non--trivial (and improved with respect to mean--field) estimate of
the dynamical transition point $\alpha_c(\beta)$, which in DWT is
given by $1/2$, independent of $\beta$. 

\begin{figure}
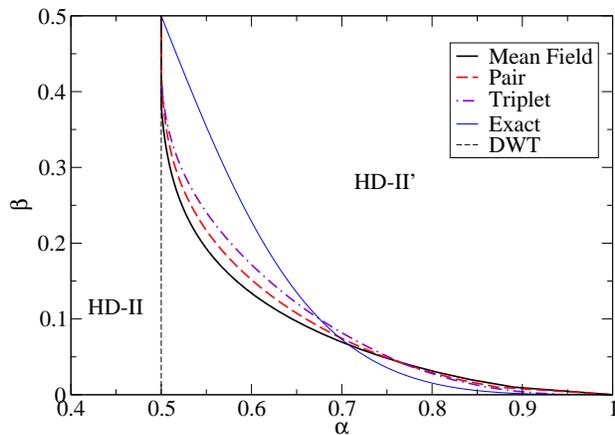

\onefigure[width=0.45\textwidth]{DynTrans.eps}
\caption{The line $\alpha_c(\beta)$ separating region HD-II from
  HD-II'. Mean field ($N = 800$, full line), pair ($N = 800$, dashed),
  triplet ($N = 800$, dot--dashed), exact (thin full line) and DWT
  (thin dashed line) results.}
\label{fig:DynTrans}
\end{figure}

The dynamical transition line $\alpha_c(\beta)$ separating region HD-II from
  HD-II' is reported in Fig.\ \ref{fig:DynTrans} and compared with
  mean--field, DWT and exact results. 

\section{Discussion}

We have shown how cluster approximations, based on suitable
factorizations of the probability distribution at a given time, can be
applied to the TASEP model with open boundaries. In particular we have
considered the pair and triplet approximations (which have been
previously applied to other models and called respectively $(2,1)$ and
$(3,2)$ approximations in \cite{nm}, and ``triplet'' and
``quintuplet'' approximations in \cite{SchweitzerBehera2015}), and
compared their results with mean--field, exact and DWT results.

We have observed that cluster approximations systematically give a
quantitative improvement with respect to mean--field results, though
retaining qualitative mean--field features. From a
computational point of view, the resources needed for both mean--field
and cluster approximations are negligible with respect to
simulations. 

We have focused in particular on the recently discovered dynamical
transition, which has been shown \cite{ProemeBlytheEvans11} to be
particularly challenging for numerical simulations. Here even simple
mean--field gives interesting results, in particular concerning the
spectrum of the relaxation matrix. It is worth observing that our
mean--field result for the dynamical transition is more accurate than
the one given by the Burgers equation approach
\cite{ProemeBlytheEvans11}. Pair and triplet approximations improve
over mean--field and give a non--trival estimate of the transition
point $\alpha_c(\beta)$, which indeed depends on $\beta$, at odds with
the DWT result. 

The results we have obtained suggest that is worthwhile to apply these
techniques to generalizations of the TASEP model for which exact
results are limited or not available, like the Partially Asymmetric
Simple Exclusion Process \cite{ChouMallickZia11} or the TASEP with
Langmuir kinetics
\cite{ParmeggianiFranoschFrey03,ParmeggianiFranoschFrey04}.

\section{Author contribution statement}

The two authors contributed equally to the paper. 



\begin{thebibliography}{0}

\bibitem{vanKampen}
  \Name{van Kampen N.}
  \Book{Stochastic Processes in Physics and Chemistry}
  \Publ{Elsevier, Amsterdam}
  \Year{2007}

\bibitem{Kawasaki}
  \Name{Kawasaki K.}
  \Book{Phase Transitions and Critical Phenomena}
  \Editor{C. Domb \and M. S. Green}
  \Vol{2}
  \Publ{Academic, London}
  \Year{1972}
  \Page{443}

\bibitem{TransportBook}
  \Name{Schadschneider A., Chowdhury D. \and Nishinari K.} 
  \Book{Stochastic Transport in Complex Systems} 
  \Publ{Elsevier, Amsterdam}
  \Year{2011}

\bibitem{nm}
  \Name{ben-Avraham D. \and K\"ohler J.}
  \REVIEW{Phys. Rev. A}{45}{1992}{8358}

\bibitem{Ito1995}
  \Name{Schrechenberg M., Schadschneider A., Nagel K. \and Ito N.}
  \REVIEW{Phys. Rev. E}{51}{1995}{2939}

\bibitem{BoccaraBook}
  \Name{Boccara N.}
  \Book{Modeling Complex Systems} 
  \Publ{Springer, New York}
  \Year{2004}

\bibitem{Knight1987}
  \Name{Gutowitz H.A., Victor J.D. \and Knight B.W.}
  \REVIEW{Physica D}{28}{1987}{18}

\bibitem{VulpianiBook}
  \Name{Crisanti A., Paladin G. \and Vulpiani A.}
  \Book{Products of Random Matrices in Statistical Physics} 
  \Publ{Springer, Berlin}
  \Year{1983}

\bibitem{AdvPhys}
  \Name{Gouyet J.-F., Plapp M., Dieterich W. \and Maass P.}
  \REVIEW{Adv. Phys.}{52}{2003}{523}

\bibitem{DeLos1}
  \Name{Petermann T. \and De Los Rios P.}
  \REVIEW{J. Theor. Biol.}{229}{2004}{1}

\bibitem{DeLos2}
  \Name{Petermann T. \and De Los Rios P.}
  \REVIEW{Phys. Rev. E}{69}{2004}{066116}

\bibitem{SchweitzerBehera2015}
  \Name{Schweitzer F. \and Behera L.}
  \REVIEW{Entropy}{17}{2015}{7658}

\bibitem{PPM0}
  \Name{Kikuchi R.}
  \REVIEW{Phys. Rev.}{124}{1961}{1682}

\bibitem{PPM01}
  \Name{Kikuchi R.}
  \REVIEW{Prog. Theor. Phys. Suppl.}{35}{1966}{1}

\bibitem{PPM1}
  \Name{Ishii T.}
  \REVIEW{Prog. Theor. Phys. Suppl.}{115}{1994}{243}

\bibitem{PPM2}
  \Name{Ducastelle F.}
  \REVIEW{Prog. Theor. Phys. Suppl.}{115}{1994}{255}

\bibitem{PPM3}
  \Name{Wada K. \and Kaburagi M.}
  \REVIEW{Prog. Theor. Phys. Suppl.}{115}{1994}{273}

\bibitem{CVM}
  \Name{Kikuchi R.}
  \REVIEW{Phys. Rev.}{81}{1951}{988}

\bibitem{An}
  \Name{An G.}
  \REVIEW{J. Stat. Phys.}{52}{1988}{727}

\bibitem{MyRev}
  \Name{Pelizzola A.}
  \REVIEW{J. Phys. A: Math. Gen.}{38}{2005}{R309}

\bibitem{ZM-JSTAT}
  \Name{Zamparo M. \and Pelizzola A.}
  \REVIEW{J. Stat. Mech.}{}{2006}{P12009}

\bibitem{Mezard}
  \Name{M\'ezard M. \and Montanari A.}
  \Book{Information, Physics and Computation}
  \Publ{Oxford University Press}
  \Year{2009}

\bibitem{DynCavity00}
  \Name{Neri I. \and Boll\'e D.}
  \REVIEW{J. Stat. Mech.}{}{2009}{P08009}

\bibitem{DynCavity0}
  \Name{Kanoria Y. \and Montanari A.}
  \REVIEW{Ann. Appl. Prob.}{21}{2011}{1694}

\bibitem{DynCavity1}
  \Name{Aurell E. \and Mahmoudi H.}
  \REVIEW{J. Stat. Mech.}{}{2011}{P04014}

\bibitem{DynCavity2}
  \Name{Aurell E. \and Mahmoudi H.}
  \REVIEW{Comm. Theor. Phys.}{56}{2011}{157}

\bibitem{DynCavity3}
  \Name{Aurell E. \and Mahmoudi H.}
  \REVIEW{Phys. Rev. E}{85}{2012}{031119}

\bibitem{DMP1}
  \Name{Lokhov A.Y., M\'ezard M. \and Zdeborov\'a L.}
  \REVIEW{Phys. Rev. E}{91}{2015}{012811}

\bibitem{DMP2}
  \Name{Del Ferraro G. \and Aurell E.}
  \REVIEW{Phys. Rev. E}{92}{2015}{010102}

\bibitem{Moore2015}
  \Name{Shrestha M., Scarpino S.V. \and Moore C.}
  \REVIEW{Phys. Rev. E}{92}{2015}{022821}

\bibitem{Pelizzola}
  \Name{Pelizzola A.}
  \REVIEW{Eur. Phys. J. B}{86}{2013}{120}

\bibitem{RicciTersenghi}
  \Name{Dominguez E., Del Ferraro G. \and Ricci-Tersenghi F.}
  \REVIEW{J. Stat. Mech.}{}{2017}{P033303}

\bibitem{Pretti-SIS}
  \Name{Pelizzola A. \and Pretti M.}
  \REVIEW{arXiv:1702.06822 [cond-mat.stat-mech]}{}{2017}{}

\bibitem{ChouMallickZia11}
  \Name{Chou T., Mallick K. \and Zia R.K.P.}
  \REVIEW{Rep. Prog. Phys.}{74}{2011}{116601}

\bibitem{MacDonaldGibbsPipkin68}
  \Name{MacDonald C.T., Gibbs J.H. \and Pipkin A.C.}
  \REVIEW{Biopolymers}{6}{1968}{1}

\bibitem{Krug91}
  \Name{Krug J.}
  \REVIEW{Phys. Rev. Lett.}{67}{1991}{1882}

\bibitem{DerridaDomanyMukamel92}
  \Name{Derrida B., Domany E. \and Mukamel D.}
  \REVIEW{J. Stat. Phys.}{69}{1992}{667}

\bibitem{SchutzDomany93}
  \Name{Sch\"utz G. \and Domany E.}
  \REVIEW{J. Stat. Phys.}{72}{1993}{277}

\bibitem{Derrida-etal93}
  \Name{Derrida B., Evans M.R., Hakim V. \and Pasquier V.}
  \REVIEW{J. Phys. A: Math. Gen.}{26}{1993}{1493}

\bibitem{Derrida98}
  \Name{Derrida B.}
  \REVIEW{Phys. Rep.}{301}{1998}{65}

\bibitem{deGierEssler05}
  \Name{deGier J. \and Essler F.H.L.}
  \REVIEW{Phys. Rev. Lett.}{95}{2005}{240601}

\bibitem{deGierEssler08}
  \Name{deGier J. \and Essler F.H.L.}
  \REVIEW{J. Phys. A: Math. Theor.}{41}{2008}{485002}

\bibitem{ProemeBlytheEvans11}
  \Name{Proeme A., Blythe R.A. \and Evans M.R.}
  \REVIEW{J. Phys. A: Math. Theor.}{44}{2011}{035003}

\bibitem{Smith}
  \Name{Smith G.D.} 
  \Book{Numerical Solution of Partial Differential Equations} 
  \Publ{Clarendon Press, Oxford}
  \Year{1978}

\bibitem{DWT1}
  \Name{Dudzi\'{n}ski M. and Sch\"utz G.M.}
  \REVIEW{J. Phys. A: Math. Gen.}{33}{2000}{8351}

\bibitem{DWT2}
  \Name{Nagy Z., Appert C. and Santen L.}
  \REVIEW{J. Stat. Phys.}{109}{2002}{623}

\bibitem{GoverBarnett}
  \Name{Gover M.J.C. \and Barnett S.}
  \REVIEW{IMA J. Numer. Anal.}{5}{1985}{101}

\bibitem{ParmeggianiFranoschFrey03}
  \Name{Parmeggiani A., Franosch T. \and Frey E.}
  \REVIEW{Phys. Rev. Lett.}{90}{2003}{086601}

\bibitem{ParmeggianiFranoschFrey04}
  \Name{Parmeggiani A., Franosch T. \and Frey E.}
  \REVIEW{Phys. Rev. E}{70}{2004}{046101}









\end{thebibliography}
\end{document}